# Room temperature wavelike exciton transport in a van der Waals superatomic semiconductor

*Jakhangirkhodja A. Tulyagankhodjaev[1], Petra Shih[1†], Jessica Yu[1†], Jake C. Russell[1], Daniel G. Chica[1], Michelle E. Reynoso[1], Haowen Su[1], Athena C. Stenor[1], Xavier Roy[1], Timothy C. Berkelbach[1], Milan Delor[1*]*

[1]Department of Chemistry, Columbia University; New York, NY 10027, USA
[†]These authors contributed equally.
*Corresponding author. Email: milan.delor@columbia.edu

**Abstract:** The transport of energy and information in semiconductors is limited by scattering between electronic carriers and lattice phonons, resulting in diffusive and lossy transport that curtails all semiconductor technologies. Using $Re_6Se_8Cl_2$, a van der Waals (vdW) superatomic semiconductor, we demonstrate the formation of acoustic exciton-polarons, an electronic quasiparticle shielded from phonon scattering. We directly image polaron transport in $Re_6Se_8Cl_2$ at room temperature and reveal quasi-ballistic, wavelike propagation sustained for nanoseconds and several microns. Shielded polaron transport leads to electronic energy propagation orders of magnitude greater than in other vdW semiconductors, exceeding even silicon over nanoseconds. We propose that, counterintuitively, quasi-flat electronic bands and strong exciton–acoustic phonon coupling are together responsible for the remarkable transport properties of $Re_6Se_8Cl_2$, establishing a new path to ballistic room-temperature semiconductors.

**Main text:**

Semiconductor technologies rely on transporting energy and information carriers, often in the form of electrons or excitons (bound electron-hole pairs), from source to target. At room temperature, these carriers rapidly scatter with lattice vibrations (phonons) on nanometer and femtosecond scales. Scattering leads to electronic energy dissipation, joule heating and loss of phase coherence and directionality, imposing strict speed and efficiency limits on all semiconductor technologies. Breaking through these limits requires semiconductors that sustain ballistic (scatter-free), wavelike flow of energy over macroscopic distances at room temperature, a long sought goal that would enable ballistic transistors (*1*), low-loss energy harvesting and wave-based information technologies (*2*).

Here, we demonstrate macroscopic, wavelike exciton flow at room temperature in the van der Waals (vdW) superatomic material $Re_6Se_8Cl_2$ (Fig. 1A). $Re_6Se_8Cl_2$ is a semiconductor with an indirect bandgap of 1.6 eV and an exciton binding energy of ~100 meV (*3*). Its superatom building blocks consist of $Re_6$ octahedra enclosed in $Se_8$ cubes. Each $Re_6Se_8$ unit is covalently bonded to four neighbors to form a two-dimensional (2D) pseudo-square lattice capped by Cl atoms at the apical positions. The $Re_6Se_8Cl_2$ layers stack out of plane to create a bulk vdW crystal with weak interlayer electronic coupling (*4*). The crystal can be exfoliated to the monolayer limit, advantageous for integration in gated devices (*3*, *5*, *6*). $Re_6Se_8Cl_2$ exhibits relatively weak inter-cluster electronic coupling, as evidenced by the electronic band structure (Fig. 1B) (*3*, *7*). Strong coupling of electrons to inter-cluster optical phonons (*8*) leads to further band flattening at room temperature (*7*) and has been implicated in the emergence of superconductivity in this material and related classes (*6*, *9*). In this work, we demonstrate that these quasi-flat electronic bands, in combination with strong coupling to acoustic phonons, lead to the formation of acoustic exciton-polarons (Fig. 1C), quasiparticles of excitons bound to an acoustic lattice deformation. Through direct imaging of polaron propagation, we reveal that they are shielded from lattice scattering, leading to quasi-ballistic transport over several microns at room temperature, currently limited only by crystal size. Our observations challenge the common notion that strong electronic coupling is required for long-range transport.

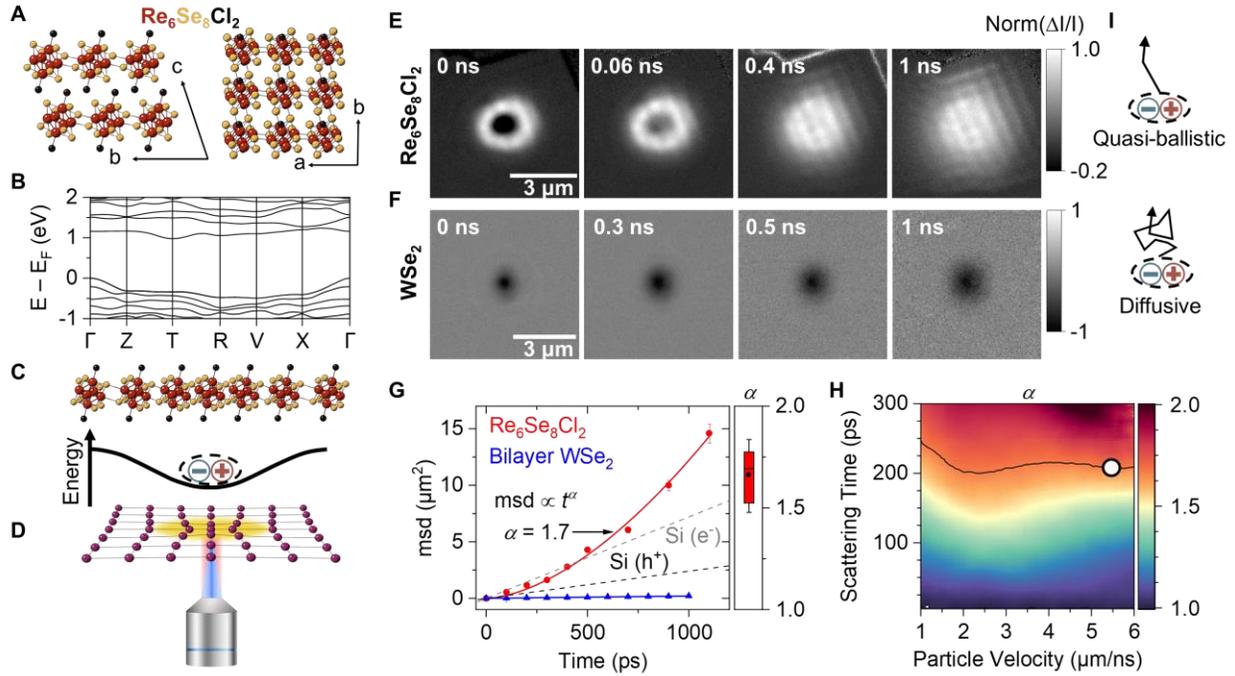

**Figure 1. Imaging exciton transport in $Re_6Se_8Cl_2$**. (A) Crystal structure of $Re_6Se_8Cl_2$. (B) Band structure of $Re_6Se_8Cl_2$ calculated at the DFT/PBE level. (C) Formation of acoustic polarons via a deformation potential interaction. (D) Schematic for optical far-field imaging of polaron transport (details in the text and figs. S1–S3). (E) stroboSCAT time series displaying exciton (dark contrast) and exciton-polaron (bright contrast) propagation in $Re_6Se_8Cl_2$. (F) Exciton propagation in bilayer $WSe_2$ on glass. (G) Mean squared displacement (msd) of exciton-polarons in $Re_6Se_8Cl_2$ (red), excitons in $WSe_2$ (blue), and charge carriers in Si (grey and black). Error bars are 1 standard deviation. Only $Re_6Se_8Cl_2$ displays superlinear behavior, indicating superdiffusive transport characterized by the exponent $\alpha$. The box plot shows the spread of $\alpha$ values across 11 different datasets, indicating mean and median values of 1.67 and 1.7, respectively. (H) Monte Carlo simulation of $\alpha$ for different particle velocities ($v$) and scattering times ($\tau$) for our experimental configuration. The circle corresponds to simulation parameters of $v = 5.5$ km/s and $\tau = 215$ ps that reproduce the experimental msd. The black contour traces $\alpha = 1.67$. (I) Illustration of quasi-ballistic motion of polarons in $Re_6Se_8Cl_2$, compared to diffusive motion for excitons in $WSe_2$ and other semiconductors.

**Exciton transport in $Re_6Se_8Cl_2$ is quasi-ballistic.** We directly image exciton transport in single-crystal $Re_6Se_8Cl_2$ using ultrafast stroboscopic scattering microscopy (*10–12*) (stroboSCAT, Fig. 1D and figs. S1–S3). An above-gap, diffraction-limited visible pump generates excitons, and then a backscattering widefield probe (1.55 eV) slightly below the electronic bandgap spatially resolves how the excitons modify the local polarizability of the material. By varying the pump-probe time delay, we spatiotemporally track the evolution of photoexcitations in an all-optical, non-invasive and contact-free measurement. Figure 1E and movie S1 display representative stroboSCAT data obtained in a 60 nm thick $Re_6Se_8Cl_2$ flake prepared by mechanical exfoliation (*13*). Two key features emerge from this data: First, the initial negative (dark) stroboSCAT contrast turns to positive (bright) contrast on a few-picosecond timescale, which we show below represents a transition from a bare exciton to an exciton-polaron. Second, the exciton-polaron propagates several microns to the edge of the flake in less than a nanosecond. This remarkably fast and long-range transport differs starkly from exciton transport in other molecular or 2D semiconductors (Table S1). For comparison, Figure 1F displays stroboSCAT data of exciton transport in the archetypal vdW semiconductor $WSe_2$ (bilayer flake on glass — see fig. S4 for monolayer and bulk data), exhibiting much slower and shorter-range transport. These results are counterintuitive, since the effective mass of excitons in $WSe_2$ is much smaller than in $Re_6Se_8Cl_2$ (Table S1).

To quantify and rationalize the remarkable transport properties of $Re_6Se_8Cl_2$, we plot the mean squared displacement of the photoexcited population profile observed in stroboSCAT as msd $= \sigma^2(t) - \sigma^2(0)$, where $\sigma$ is the Gaussian width of the population density profile at time delay $t$ (*13*). Figure 1G compares the msd for exciton-polarons in $Re_6Se_8Cl_2$ against the msd for excitons in bilayer $WSe_2$ and charge carriers in intrinsic monocrystalline Si (*14*), which exhibit some of the best transport among 2D and 3D semiconductors, respectively. We find that the msd at 1.1 ns in $Re_6Se_8Cl_2$ is 23 times that in bulk $WSe_2$, 65 times that in bilayer $WSe_2$, 120 times that in monolayer $WSe_2$, and remarkably almost twice that of electrons in intrinsic Si and the recently-reported cubic boron arsenide (*15*, *16*) (Table S1). Based on the 11 ns polaron lifetime (fig. S5), we estimate that the polaron propagation length in $Re_6Se_8Cl_2$ would exceed 25 μm in the absence of crystal boundaries.

The superlinear behavior of the msd for $Re_6Se_8Cl_2$ differs from the linear behavior in Si and $WSe_2$. We fit the msd to a power law, msd $\propto t^\alpha$ (*11*, *12*, *17*). In the limit of diffusive transport, where scattering lengths are much shorter than the propagation length, $\alpha = 1$. This regime, exemplified by the linear msd for Si and $WSe_2$ in Figure 1F, is observed in virtually all semiconductors beyond the first few femtoseconds following photoexcitation (*18*). In the limit of coherent, ballistic transport (no scattering), $\alpha = 2$, i.e. distance is proportional to time with a slope that defines the velocity. In $Re_6Se_8Cl_2$, we observe quasi-ballistic transport ($\alpha = 1.67 \pm 0.13$, Fig. 1G) sustained for nanoseconds, until the flake edge is reached. Plotting the same data as distance vs. time provides an effective propagation velocity of 2.3 km/s (fig. S6). Monte Carlo simulations reproducing the observed msd yield an exciton–lattice scattering time of 215 ps, indicating an extraordinary mean free path between scattering events exceeding 1 μm for exciton-polarons in $Re_6Se_8Cl_2$ (Fig. 1H; see figs. S7–S8 for simulation details and alternative models). These findings reveal that the mechanism for fast and long-range transport in $Re_6Se_8Cl_2$ is efficient shielding from scattering, amply compensating for the large effective mass (and thus low intrinsic velocity) of the polaron. These results are reproducible in multiple flakes of different thicknesses, for both above-gap and band-edge pump excitation, for pump temporal pulsewidths spanning 50 fs–50 ps, and across a range of fluences explored in detail below, indicating that hot carrier transport (*19*), phonon winds (*20*, *21*), nonlinear recombination (*22*), thermal gradients (*23*), or strain waves (*24*–*27*) are not responsible for the observed behavior (fig. S9–S11, Table S2).

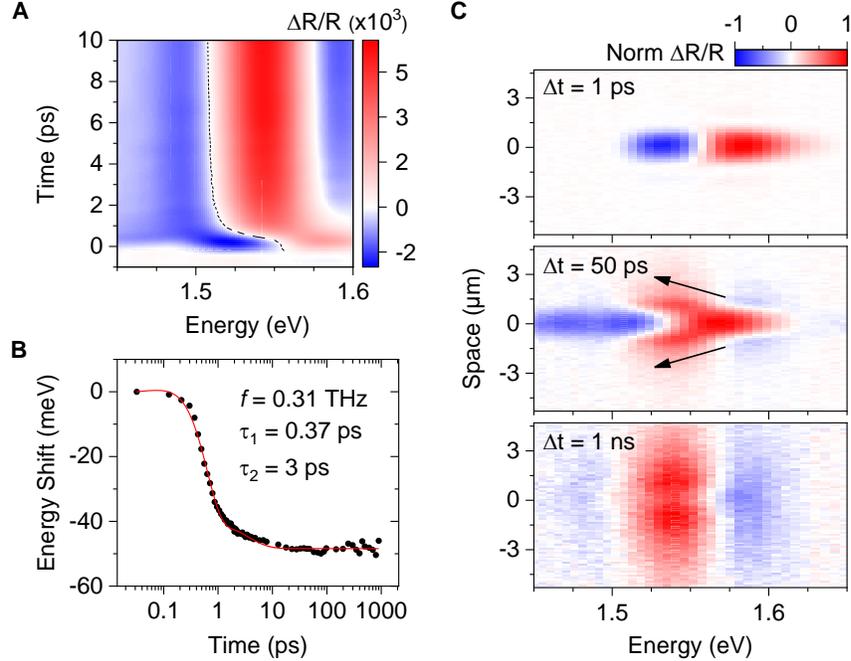

**Figure 2. Spatio-energetic tracking of polaron formation in Re$_6$Se$_8$Cl$_2$.** (A) Transient reflectance spectra near the band edge following 2.41 eV pump excitation at an initial exciton density of 5 x 10$^{17}$/cm$^3$. (B) Trace of the redshift highlighted with a dashed line in panel A. The red line is a fit using a damped oscillator model with a frequency of 0.31 THz and damping time of 0.37 ps, combined with a 3 ps exponential decay. (C) Spacetime-transient reflection spectra following 2.41 eV pump excitation at an initial exciton density of 8 x 10$^{18}$/cm$^3$ for three time delays. The correlated redshift and transport dynamics are emphasized with arrows in the middle panel. Supplementary Movies S2, S3 display these coupled spatio-energetic dynamics for different excitation fluences.

**Excitons in Re$_6$Se$_8$Cl$_2$ form exciton-polarons.** To confirm that polaron formation is responsible for the observed transport behavior, we track the energetic evolution of excitons following photoexcitation and correlate these dynamics to optical transport measurements. Figure 2A displays transient reflectance spectra in the region of the semiconductor band edge, exhibiting a bleach around 1.57 eV and a photoinduced absorption ~90 meV higher in energy. The primary dynamic evolution observed is a spectral redshift on a 1.5 ps timescale that convolves all peaks. Identical dynamics are observed for near band-edge excitation (fig. S9), ruling out the possibility that electronic thermalization is responsible for the redshift. Tracking the zero-crossing of the transient reflectance profile (dashed line in Fig 2A) provides a handle on redshift kinetics, plotted in Figure 2B. We observe an overall 48 meV redshift with an evolution resembling a strongly damped oscillator (red line in Fig. 2B). These dynamics echo those previously observed for (exciton- or large) polaron formation (*28–31*), wherein energetic stabilization occurs over a single vibrational period of the associated lattice deformation. This spectral evolution is responsible for the switch from dark to bright contrast in stroboSCAT in Fig. 1E: the probe at 1.55 eV is initially pre-resonant with the exciton transition (dark contrast), but switches to resonant with the induced polaron absorption following the redshift, generating a bright contrast (additional stroboSCAT datasets at different probe wavelengths are displayed in fig. S12). Below, we discuss measurements of correlated spatio-energetic dynamics, fluence-dependent redshift dynamics and population saturation that cement our assignment of these redshift dynamics to polaron formation.

Polarons should exhibit correlated spatial and spectral dynamics, since energetic stabilization associated with polaron formation results in a modification of transport properties. To directly image these correlated dynamics and further support the correspondence between the redshift timescale and the polaron formation timescale, we develop a new approach capable of simultaneously resolving spectral and spatial evolution by merging stroboSCAT with transient reflectance microscopy (fig. S2). Figure 2C displays spatially-resolved transient spectra at different pump-probe time delays (see also movies S2–S3). At early times, the transient reflectance signal associated with excitons is concentrated around the pump excitation location at 0 μm. Between 4 and 50 ps, a V-shaped pattern emerges (highlighted with arrows in Fig. 2C), indicating that the spectral redshift associated with polaron formation is correlated to transport away from the excitation location. By 1 ns, only the redshifted polaron spectral signature remains, showing propagation over several microns. The correlated spatio-energetic dynamics are a clear indication that excitons in $Re_6Se_8Cl_2$ become substantially mobile only after they have formed exciton-polarons; bare excitons are effectively immobile, whereas polarons propagate over microns. The observed transport dynamics also rule out traditional exciton self-trapping (small polarons), which would reduce the exciton mobility upon polaron formation (*32*). These results illustrate the power of our correlative approach for understanding how polaron formation affects transport.

Finally, we confirm the formation of polarons in $Re_6Se_8Cl_2$ and experimentally infer their size by determining the density at which they begin to interact. At high densities, lattice deformations compete to displace the same atoms, resulting in a diminished ability to form polarons (Fig. 3A) (*32*). stroboSCAT data displayed in Fig. 3B shows that as the photoexcitation density increases, the bare exciton signal (dark contrast) begins to dominate over the polaron signal (bright contrast), indicating that polaron formation is suppressed in regions of high excitation density. Figure 3C compares the polaron population (red trace) and bare exciton population (black trace) as a function of excitation fluence (analysis in fig. S13). We observe clear saturation of the polaron population at exciton densities between $0.45 \times 10^{18}/cm^3$ and $1.8 \times 10^{18}/cm^3$ (highlighted with a blue rectangle). This behavior signals that polarons are overlapping in space, analogous to a Mott transition that prevents further polaron formation (*33*). The onset of polaron saturation is also reflected in a transition from superdiffusive to almost diffusive transport (Fig. 3D), which we attribute to polaron–polaron scattering above the saturation density. The suppression of exciton-to-polaron conversion above saturation is most evident in the spectral dynamics (Fig. 3E and fig. S14), where the redshift time associated with exciton stabilization increases from ~1.5 ps to hundreds of ps. We reproduce these spectral dynamics (right panels of Fig. 3E) with a saturation model accounting for a kinetic blockade and polaron transport away from the excitation area (figs. S14–S15). The polaron interaction radius associated with the observed critical density range of $(0.45–1.8) \times 10^{18}/cm^3$ is 2.1–3.4 nm within the Mott criterion (*34*), corresponding to 3–5 unit cells in $Re_6Se_8Cl_2$.

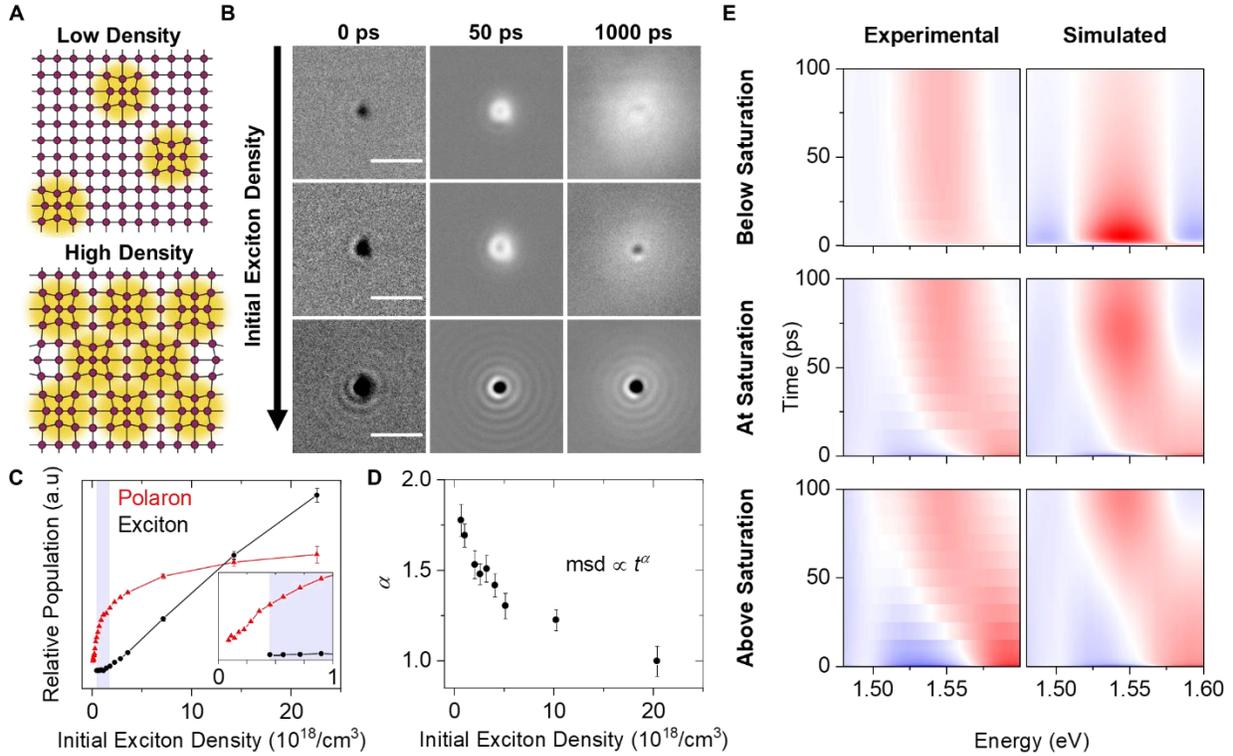

**Figure 3. Determining polaron size.** (A) Depiction of polaron blockade at high densities. (B) stroboSCAT images at pump/probe energies of 2.41 eV/1.55 eV at initial exciton densities ranging from (1–20) x $10^{18}$ /cm$^3$ in a 350 nm thick flake. Bare excitons are associated with dark contrast, whereas polarons are associated with bright contrast. The rings observed at high exciton densities are diffraction rings. All scale bars are 3 μm. (C) Relative populations of polarons and bare excitons at a pump-probe time delay of 5 ps as a function of initial exciton density, indicating saturation of the polaron population. The blue shaded region for exciton densities between 0.45 x $10^{18}$ and 1.8 x $10^{18}$ /cm$^3$ indicates the range of critical polaron overlap density. (D) msd exponent, $\alpha$, as a function of initial exciton density. Error bars are 1 standard deviation. (E) Experimental (left) and simulated (right) transient reflectance spectra taken at the center of the focused pump excitation (panel B) for different initial exciton densities. Simulations are based on a saturation model accounting for exciton to polaron conversion and polaron transport away from the excitation spot (*13*).

**Acoustic polarons are responsible for wavelike transport.** Large polarons, known to form in materials such as lead halide perovskite semiconductors, have been suggested to partially shield carrier-lattice scattering (*35*). Nevertheless, experiments consistently demonstrate a diffusive transport regime (*10*, *18*, *19*, *36*), with sub-100 fs scattering times that indicate insufficient shielding to switch into the much-desired macroscopic ballistic transport regime. In contrast, the sustained quasi-ballistic behavior observed in Re$_6$Se$_8$Cl$_2$ is reminiscent of acoustic polarons, which can form in low-dimensional materials and were theoretically invoked to rationalize the transport properties of polydiacetylene in one dimension (*37*, *38*).

The formation of acoustic polarons is rare, and our observation of micron-scale exciton mean free paths at room temperature is unprecedented. To rationalize this striking behavior in Re$_6$Se$_8$Cl$_2$, we employ an approximate strong-coupling theory that describes an exciton of mass $m$ coupled to acoustic phonons with a strength quantified by the deformation potential $D$ in two dimensions. The acoustic phonons derive from superatoms of mass $M$ with inter-cluster vibrational frequency $\Omega$. In two dimensions, the energy to form a circular polaron with radius $a$ and area $\pi a^2$ is (*39*)

$$E = \frac{\hbar^2}{2ma^2} + \frac{1}{2}M\Omega^2\Delta^2\pi a^2 - D\Delta \qquad (1)$$

where $\Delta$ is the dimensionless lattice displacement. In this simple picture, the polaron is only bound if the electron-phonon interaction outweighs the energetic penalties associated with exciton localization and lattice deformation. Minimizing equation (1) with respect to the lattice displacement $\Delta$ gives the existence criterion

$$\lambda > \lambda_c, \qquad \lambda = \frac{mD^2}{2\hbar^2 M\Omega^2} = \frac{D^2}{4JMs^2}, \qquad \lambda_c = \pi/2 \approx 1.6 \qquad (2)$$

where $\lambda$ is a dimensionless measure of the exciton–acoustic phonon coupling strength, and in the second equality we have introduced the exciton transfer integral $J$ and the speed of sound $s = \Omega a$. Applying density functional theory to $Re_6Se_8Cl_2$, we calculate $D = 4.4$ eV. Taking into account the experimentally-inferred electronic band flattening due to inter-cluster optical phonons, which increases the exciton effective mass from 1.9 $m_e$ at 0 K to 60 $m_e$ at 300 K (*7*, *13*), we calculate a very small $J$ of 1.5 meV at 300 K. When combined with the other material parameters (*13*), we find that $\lambda = 7 > \lambda_c$, predicting a strongly bound polaron. Figure 4A plots the coupling strength $\lambda$ for $Re_6Se_8Cl_2$, monolayer $WSe_2$, crystalline pentacene and 2D organic-inorganic halide perovskites. $Re_6Se_8Cl_2$ has a coupling strength $\lambda$ that is 10–1000 times larger than that of the other materials and is thus the only one predicted to exhibit bound acoustic polarons according to the criterion in equation (2). Within this theory, the key parameters setting $Re_6Se_8Cl_2$ apart from other two- and three-dimensional semiconductors is the combination of a quasi flat-band electronic structure at room temperature (small $J$) and strong exciton–acoustic phonon interactions (large $D$), yielding a large coupling strength $\lambda$ and associated strongly bound acoustic polarons. The polaron binding energy is reduced with decreasing temperature due to the increase in the transfer integral $J$ (*7*, *13*). The estimate in equation (2) predicts that the polaron is not bound below ~175 K. Temperature-dependent stroboSCAT experiments (fig. S16) display a dramatic reduction in msd below ~150 K, lending support to our central hypothesis and theory of acoustic polarons. We emphasize that equation (2) only provides a qualitative criterion for polaron formation and a detailed understanding of the polaron stability and lifetime at finite temperature requires a more complete theoretical treatment (*13*).

Rationalizing the quasi-ballistic dynamics of acoustic polarons requires a more sophisticated quantum mechanical treatment. We generalize the stationary polaron description above and propose a variational wavefunction for the moving polaron defined by its average crystal momentum (*13*). Energy minimization produces the polaron dispersion shown in Fig. 4B. Near the band bottom, we extract a large effective polaron mass $m^* \approx 200\ m_e$, a substantial increase from the bare exciton mass of 60 $m_e$ at 300 K. More significantly, at higher momenta, the polaron inherits the linear dispersion of acoustic phonons — a renormalization evocative of light-matter hybridization to form polaritons (*40*). The linear dispersion of acoustic polarons with a slope below that of other acoustic phonons implies weak scattering, since there are no dissipation channels that conserve both energy and momentum (*38*). The polaron is thus predicted to move quasi-ballistically at a speed proportional to the speed of sound of the lattice, consistent with our experimental observations.

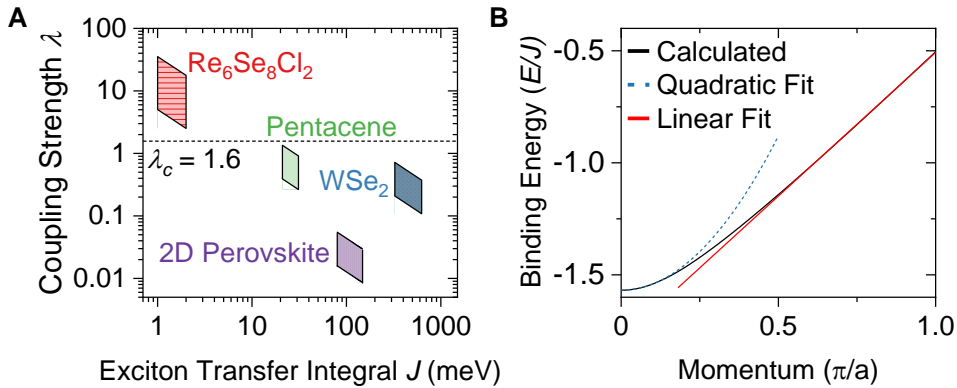

**Figure 4. Energy of an acoustic polaron**. (A) The coupling strength $\lambda$ for different systems, presented as a function of their exciton transfer integral $J$ (material properties collected in Table S3). The boxed regions represent the upper and lower bounds calculated for a range of deformation potentials and exciton transfer integrals within 30% of reported data. The critical value of $\lambda_c = 1.6$ is plotted as a dashed line. (B) Calculated polaron dispersion. The band effective mass is estimated by a parabolic fit at the minimum (dashed blue line). The red solid line emphasizes the linearity of the dispersion for higher values of momentum.

We have observed a new transport regime mediated by acoustic exciton-polarons in the vdW superatomic semiconductor $Re_6Se_8Cl_2$. Polaron formation shields excitons from scattering with lattice phonons, resulting in quasi-ballistic electronic energy flow over several microns within a nanosecond at room temperature. We reveal an extraordinary exciton mean free path of ~1 μm, suggesting the possibility of ballistic excitonic transistors. Our discovery of record-breaking transport in a material with weak electronic dispersion is astounding, providing an alternative to the current paradigm of increasing electronic conjugation to improve transport. Indeed, our model for 2D acoustic polarons suggests that quasi-flat electronic bands and strong electron-phonon interactions can counterintuitively result in exceptional electronic transport. Beyond 2D superatomic materials such as $Re_6Se_8Cl_2$ and the recently reported graphullerene (*41*), moiré superlattices of 2D semiconductors may provide an interesting testing ground for acoustic polarons. Their superlattice potentials enable tuning electronic bands (*42*, *43*) to achieve values of $J$ down to ~0. Combined with their strong deformation potentials (*44*), these flat bands could yield acoustic polarons with tunable transport properties across a large temperature range. Generalizing wavelike, ultralong-range electronic energy flow in 2D materials could herald a new era of essentially lossless nanoelectronics.

**Acknowledgements:**
We thank Professors Louis E. Brus, Xiaoyang Zhu, Colin Nuckolls and David R. Reichman for helpful discussions, and Ding Xu and James Baxter for technical help with measurements and analysis.

**Funding:** This material is primarily based upon work supported by the National Science Foundation (NSF) through the Columbia MRSEC on Precision-Assembled Quantum Materials (PAQM) (DMR-2011738) (MD, XR, TB) and by the Air Force Office of Scientific Research (AFOSR) under grant number FA9550-22-1-0389 (MD, XR). stroboSCAT instrument development was supported by the NSF under grant number DMR-2115625 (MD). MD acknowledges support from the Arnold and Mabel Beckman Foundation through a Beckman Young Investigator award. XR and JY acknowledge support from the National Science Foundation CAREER Award DMR-1751949. JAT was supported by the NSF Graduate Research Fellowship. JCR was supported by the Department of Defense National Defense Science and Engineering Graduate Fellowship.

**Author contributions:** JAT and MD conceived and designed the experiments. JAT performed, analyzed, and simulated stroboSCAT and STR experiments with assistance from ACS. JY, JCR, DGC and XR synthesized and characterized $Re_6Se_8Cl_2$ single crystals. JY performed and analyzed heat capacity measurements. JAT, MER, JY and HS prepared exfoliated samples. PS and TCB developed the theory and performed DFT calculations and polaron dispersion calculations. MD supervised the project. JAT, PS, JY, XR, TCB and MD wrote the manuscript, with input from all authors.

**Competing interests:** Authors declare that they have no competing interests.

**Data and materials availability:** All data are available in the main text or the supplementary materials. Raw files are available from the corresponding author upon reasonable request.

# Supplementary Materials for

## Room temperature wavelike exciton transport in a van der Waals superatomic semiconductor

Jakhangirkhodja A. Tulyagankhodjaev, Petra Shih, Jessica Yu, Jake C. Russell, Daniel G. Chica, Michelle E. Reynoso, Haowen Su, Athena C. Stenor, Xavier Roy, Timothy C. Berkelbach, Milan Delor

Correspondence to: milan.delor@columbia.edu

# 1 Materials and Methods

## 1.1 Synthesis, exfoliation, and characterization of $Re_6Se_8Cl_2$

$Re_6Se_8Cl_2$ crystals were synthesized via chemical vapor transport following previous reports (*3*, *6*). Re (330 mg, 1.77 mmol), Se (190 mg, 2.41 mmol) and $ReCl_5$ (200 mg, 0.55 mmol) were pressed into a pellet and sealed in a 30 cm quartz tube under a pressure of ~30 mtorr. The tube was heated at 1 °C/min to 1100 °C in a three-zone tube furnace and held for 3 days. To achieve single crystals, the tube was then cooled over seven hours to a temperature gradient with the hot end held at 970 °C and the cool end fluctuating between 970 °C and 925 °C for three cycles over 200 hours. The tube was further cooled at 6.3 °C/min to a gradient of 340–295 °C, and the furnace was then shut off. Excess volatile components were separated from the single crystals using a gradient of 300–25 °C, and large ~0.5 mm single crystals were recovered from the middle section of the tube. Thin $Re_6Se_8Cl_2$ flakes were subsequently prepared from single crystals by mechanical exfoliation using a polyvinyl chloride adhesive surface protective tape (ProTapes Nitto SPV224) and transferred onto a borosilicate substrate (#1.5 cover glass, 0.17 mm). Once the crystals were exfoliated onto the cover glass, atomic force microscopy (AFM) images and height profiles were acquired in PeakForce Quantitative Nanoscale Mapping in tapping mode using a Bruker Dimension FastScan AFM under ambient conditions.

## 1.2 StroboSCAT and Spacetime Transient Reflectance (STR)

Time-resolved optical measurements were carried out using two separate instruments depending on the time resolution and dynamic range needed. The first is a femtosecond system operating at 1 MHz repetition rate, with ~60 fs pulses, a temporal instrument response function (IRF) of 187 fs, and a maximum pump-probe time delay of 2.2 ns. The second is a picosecond system operating at 7.814 MHz with ~60 ps pulses, a temporal IRF of 98 ps, and a maximum pump-probe time delay up of 125 ns. All experiments are conducted at room temperature.

### 1.2.1 Femtosecond stroboSCAT and STR

For ultrafast stroboSCAT measurements (fig. S1), a 40 W Yb:KGW ultrafast regenerative amplifier (Light Conversion Carbide, 40 W, 1030 nm fundamental, 1 MHz repetition rate) seeds an optical parametric amplifier (OPA, Light Conversion, Orpheus-F) with a signal tuning range of 640–940 nm and an average pulsewidth of 60 fs. Unless otherwise stated, stroboSCAT and STR experiments use the $2^{nd}$ harmonic of the fundamental (515 nm) as the pump pulse. STR experiments using lower-energy pump pulses (e.g. for band-edge excitation) use the tunable signal output of the OPA as pump pulses. stroboSCAT probe pulses are generated using the signal output of the OPA. STR probe pulses use broadband white light (~550–980 nm) generated by focusing the fundamental into a 4 mm YAG window (EKSMA Optics, 555-7124). Dispersion of the OPA output caused by refractive optics (the microscope objective in particular) is partially pre-compensated using a pair of chirped mirrors (Venteon DCM7). For all experimental configurations, the pump pulse train is modulated at 647 Hz using an optical chopper and is collimated into a high numerical-aperture objective (Leica HC Plan Apo 63x, 1.4 NA oil immersion), resulting in a diffraction-limited excitation on the sample. The probe is sent to a double-passed, computer-controlled, mechanical delay line for control over the pump-probe time delay and is combined with the pump beam through a dichroic mirror. An $f = 250$ mm widefield lens is inserted prior to the dichroic mirror to focus the probe into the back focal plane of the objective, resulting in widefield illumination of the sample. Probe light reflected from the sample-substrate interface and scattered light from the sample are collected along the same path through the objective and are directed to a complementary metal-oxide-semiconductor camera (FLIR BFS-U3-28S5M-C with a Sony IMX 421 global shutter sensor) through an imaging lens. For stroboSCAT measurements, a bandpass filter is placed before the camera to eliminate stray light and select a narrow spectral region of the probe. Temperature-dependent stroboSCAT is performed with an identical setup, but with the sample mounted in a Montana Instruments s100 closed-loop Helium cryostat with a base temperature down to 3.5 K. The cryostat is equipped with an in-vacuum objective (Zeiss LD EC Epiplan-Neofluar 100x/0.90 DIC, NA = 0.9).

STR measurements are carried out using a homebuilt imaging spectrometer as illustrated in figures S1B and S2. The stroboSCAT image formed by a white light probe passes through a slit aperture with a width of ~500 nm (after accounting for magnification) placed in the image plane. The transmitted 'vertical image' is then collimated, dispersed horizontally using a prism (ThorLabs PS853), and focused on a CMOS array camera (FLIR BFS-U3-28S5M-C). Both the spectrometer camera and the real-space camera are triggered at twice the pump modulation rate, allowing the consecutive acquisition of images with the pump ON followed by the pump OFF. Consecutive frames are then processed according to (pump on/pump off – 1).

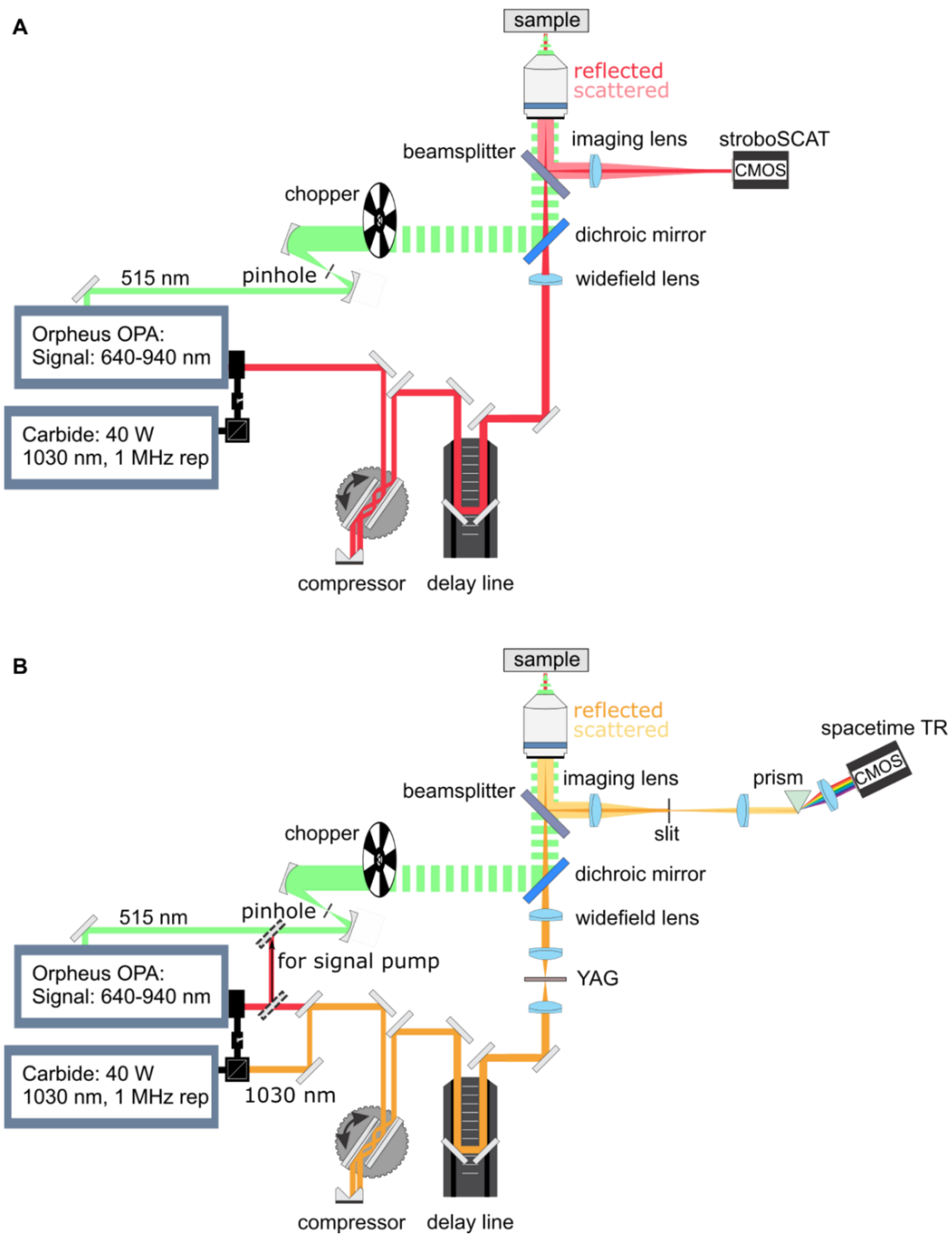

**Figure S1. Femtosecond system.** (A) Configuration for femtosecond stroboSCAT. (B) Configuration for spacetime transient reflection (STR). 1030 nm (orange) is rerouted into the original signal path. A YAG crystal generates the white light. The light is rerouted into a homebuilt spectrometer. The signal can be optionally rerouted into the 515 nm line to be used as a pump as shown by the dashed mirrors.

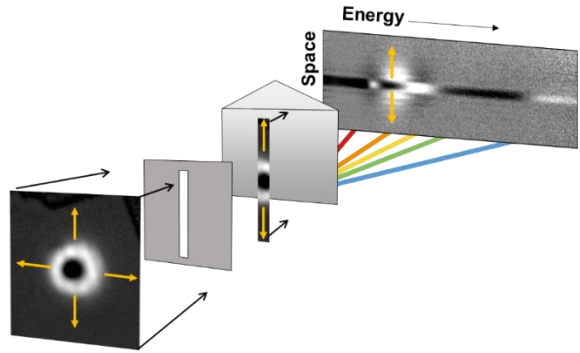

**Figure S2**. **Simplified illustration for the principle of spacetime transient reflectance.** A slit is inserted in the imaging plane. The image selected by the slit gets spectrally resolved by a spectrometer before being imaged on a 2D camera, resulting in a spatial axis along one dimension and a spectral axis along the other.

1.2.2 Picosecond stroboSCAT

For picosecond stroboSCAT (fig. S3), probe pulses are obtained from an NKT Photonics SuperK Extreme (Fianium) white light supercontinuum laser coupled into an acousto-optic tunable filter (AOTF) to select a tunable probe wavelength range from 500–900 nm with a ~30 ps pulsewidth. The pump pulses are obtained using a PicoQuant laser diode (LDH-D-C-440, 440 nm, 60 ps pulsewidth) driven by a PicoQuant laser driver (PDL-828-S "SEPIA II" equipped with a SOM 828-D oscillator). The two laser sources are synchronized by triggering the laser diode driver using a pulse train synchronization signal from the supercontinuum laser at a repetition rate of 7.814 MHz. Pump-probe time delays are controlled using the electronic delay capabilities of the diode driver, which are used to negatively delay the diode pulses with respect to the supercontinuum pulses with 20 ps resolution.

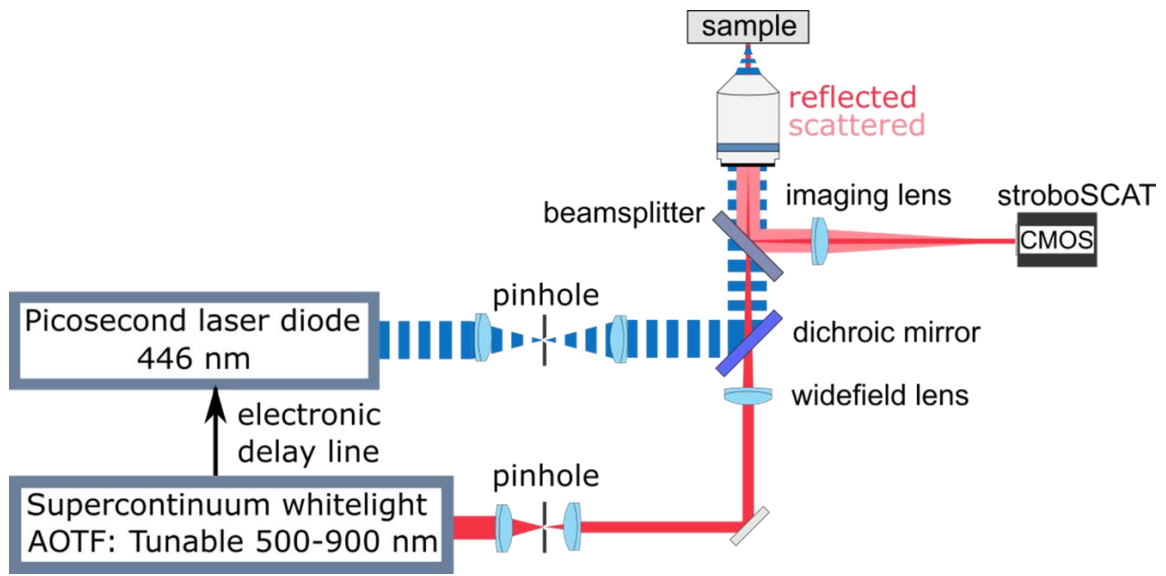

**Figure S3. Picosecond stroboSCAT system.**

## 2  Supplementary Text

| Material | Energy Type | Diffusivity/Effective Speed | Mobility ($cm^2/Vs$) | Band Effective Mass ($m_e$) |
|---|---|---|---|---|
| GaAs (45) | e⁻ | 62–232 $cm^2/s$ | 2400–9000 | 0.082–0.75 |
|  | h⁺ | 8.3–10.3 $cm^2/s$ | 320–400 | 0.57 |
| CdTe (45) | e⁻ | 27 $cm^2/s$ | 1044 | ~0.095 |
|  | h⁺ | 1.6 $cm^2/s$ | 60 | 0.12–0.81 |
| Si (45) | e⁻ | 37 $cm^2/s$ | 1450 | 0.1905–0.9163 |
|  | h⁺ | 13 $cm^2/s$ | 505 | 0.153–0.537 |
| Cubic Boron Arsenide (15, 46) | e⁻/h⁺ (ambipolar) | 41.4 $cm^2/s$ | 1600 | 0.136–1.093 |
| $CH_3NH_3PbBr_3$ (Metal Halide Perovskite) (28, 47) | e⁻ polaron | 3.9 $cm^2/s$ | 149.8 | 0.343 |
|  | h⁺ polaron | 2 $cm^2/s$ | 79.2 | 0.448 |
| Bulk $WSe_2$ [a] | Exciton | 5.7 $cm^2/s$ | -- | 0.68 |
| Monolayer $WSe_2$ [a] (48) | Exciton | 0.55 $cm^2/s$ | -- | 0.72 |
| $Re_6Se_8Cl_2$ [a, b, c] | Acoustic exciton-polarons | 65 $cm^2/s$ \| 2300 m/s | -- | 1.88 \| 203[d] |
| Pentacene (49–51) | Triplet Exciton | 3.5 x $10^{-3}$ $cm^2/s$ | -- | 3.8–12 |

[a] Diffusivity comes from a stroboSCAT measurement reported in fig. S4.

[b] To allow direct comparison against other semiconductors, the effective diffusivity and effective velocity are both reported for $Re_6Se_8Cl_2$. The effective velocity is described in the main text and section 2.2. The effective diffusivity is calculated via a linear fit from 500 to 1000 ps of Figure 1G.

[c] We report the effective mass from the zero-temperature band structure for comparison with other materials. For $Re_6Se_8Cl_2$, the effective mass (electronic dispersion) becomes much larger (flatter) at room temperature, as detailed in the main text.

[d] Exciton effective mass | Acoustic exciton-polaron effective mass

**Table S1. Comparison of effective masses and mobilities of different materials to $Re_6Se_8Cl_2$.** Materials are organized by increasing zero-temperature band effective mass (from electron, hole, or exciton bands). The dominant energy carrier is listed under energy type along with its diffusivity ($D$). For charged carriers, a mobility ($\mu$) was calculated using the Einstein relation, $\mu = eD/k_BT$, where $e$ is the free electron charge, and $T$ is the temperature. Exciton effective masses are reported within the effective mass approximation as $m_h + m_e$. Given the complexities associated with momentum and band dependence of the effective mass, a range of values between the minimum and maximum experimental values are reported (45).

## 2.1 General transport characterization

stroboSCAT and STR are both capable of directly measuring the density profile of photoexcited species across several orders of magnitude in space and time. The density profiles for these measurements are azimuthally averaged by taking the arithmetic mean of pixel values along the circumference of a circle. Gaussian functions of the form

$$\rho(r, t; \lambda) = \rho_0(t)\, \varepsilon(\lambda)\, e^{-\frac{(r-r_0)^2}{2\sigma^2(t)}}$$

are then utilized to fit to the measured density profiles along the spatial axis $r$, where $\rho_0(t)$ represents the true photoexcitation density at the center, $\varepsilon(\lambda)$ is a spectral modification parameter to account for wavelength dependent scattering/absorption cross-sections, $r_0$ marks the origin, and $\sigma(t)$ is a time dependent Gaussian width parameter.

For transport characterizations, $\rho_0(t)$ and $\varepsilon(\lambda)$ are combined into a single amplitude parameter $A(t; \lambda) = \rho_0(t)\varepsilon(\lambda)$, which becomes representative of the density. In the instance when multiple species are present, such as when the polaron blockade occurs at high fluences, additional Gaussian profiles are introduced so that the net density profile becomes $\rho(r, t; \lambda) = \rho_X + \rho_P$, where the subscripts $X$ and $P$ refer to bare excitons and exciton-polarons, respectively. Due to the large difference in width expansion between slow moving excitons and rapidly propagating polarons, double Gaussian fits are well constrained with low inter-Gaussian dependencies. All nonlinear fitting procedures use a Levenberg-Marquardt algorithm to minimize the least squares.

Standard diffusion theory predicts that the squared width of a Gaussian follows

$$\sigma^2(t) = \sigma^2(0) + 2Dt$$

where $\sigma^2(0)$ is the initial width of the Gaussian at $t = 0$ and $D$ is the diffusion constant (*12*). We refer to $\sigma^2(t) - \sigma^2(0)$ as the mean squared displacement (msd). $D$ is typically determined by a linear fit to the msd. However, as discussed in the main text, the msd can appear nonlinear. In this instance, we characterize the msd as a power law, $\sigma^2(t) - \sigma(0)^2 = At^\alpha$, where $\alpha = 2$ is classified as ballistic, $2 > \alpha \geq 1$ is considered superdiffusive, $\alpha = 1$ is diffusive and $\alpha < 1$ is subdiffusive.

It is helpful to further characterize a nonlinear msd by a linear fit to the width of the Gaussian profile $\sigma(t)$, providing a measure of an "effective" velocity. Figure S6 shows a linear fit to the function $\sigma(t) - \sigma(0) = V_{eff} t$, where $V_{eff}$ is the effective velocity. In conjunction with the polaron lifetime, $\tau_P$ (fig. S5), we use the effective velocity as an alternative definition for the propagation length/diffusion length, $L = V_{eff} \tau_P$ as opposed to the usual $L = \sqrt{2D\tau}$, due to a nonlinear msd (*12*).

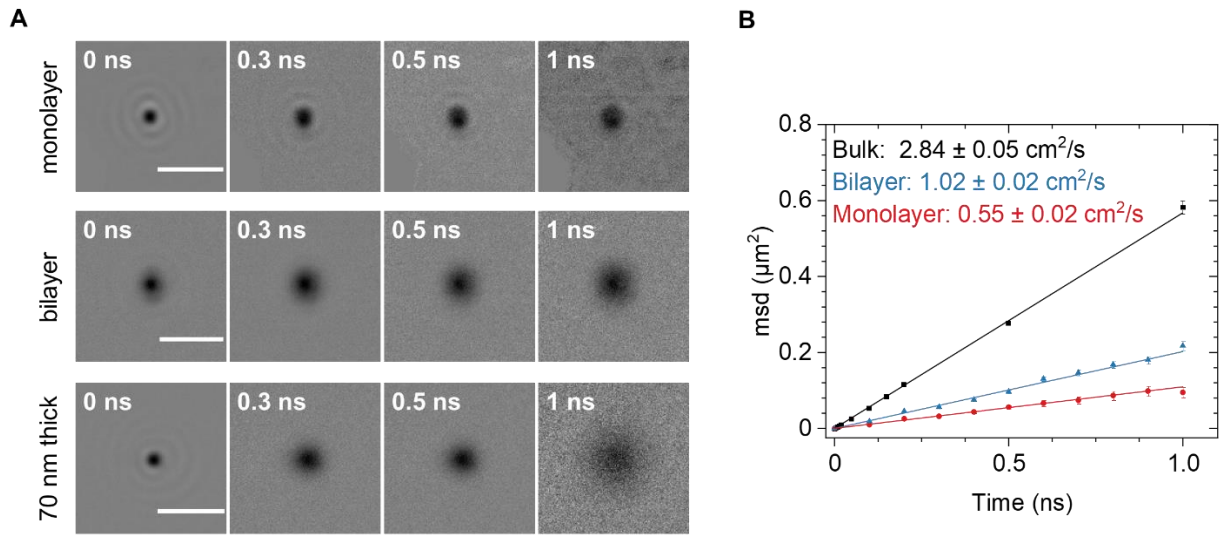

**Figure S4. Exciton transport in WSe$_2$**. (A) stroboSCAT data for monolayer, bilayer and 70 nm thick (bulk) WSe$_2$ on glass. (B) Mean squared displacement and extracted diffusion coefficients for the three datasets, consistent with previous reports (*22*, *52*). Scale bars are 3 μm.

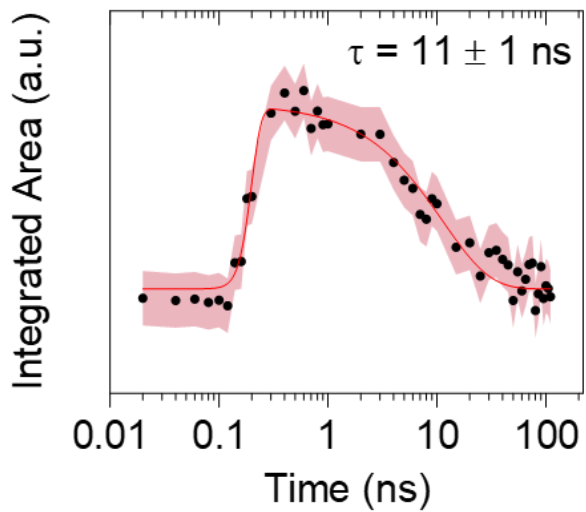

**Figure S5. Polaron lifetime in Re$_6$Se$_8$Cl$_2$.** The measurement was carried out at pre-saturation densities using picosecond stroboSCAT (fig. S3).

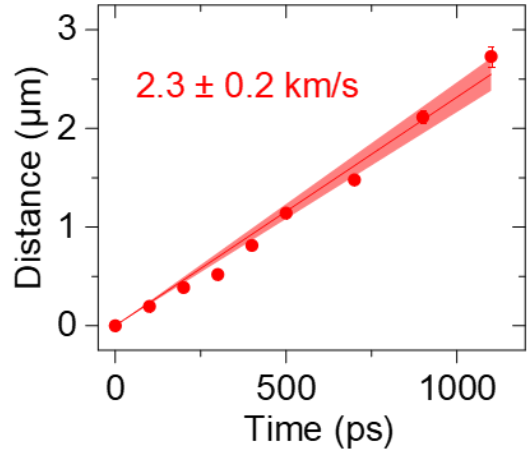

**Figure S6. Effective velocity of the polaron population profile in Re$_6$Se$_8$Cl$_2$.** The data plots $\sigma - \sigma_0$ as opposed to $\sigma^2 - \sigma_0^2$ in Fig 1G. A linear fit provides an estimated effective velocity of 2.3 ± 0.2 km/s for the population profile. The shaded red indicates the 95% confidence band in the fit.

2.2  Monte Carlo simulations of superdiffusive transport

Monte Carlo (MC) simulations were carried out to model superdiffusive transport and extract the scattering time and mean free path of polarons. Although these parameters may be extracted from a Langevin transport analysis of the msd, MC simulations allow us to account for a finite experimental field of view (11.6 μm x 11.6 μm) and observation time ($T_{obs} = 1100$ ps) over which the signal-to-noise ratio (SNR) of our measurements is sufficient to reliably fit Gaussian density profiles. The simulations generate the trajectories of 10,000 non-interacting particles that scatter with a probability $dt/\tau$ every iteration, where $dt = 1$ ps is the step size and $\tau$ is the scattering time. The particles move with a velocity $v$ in 2 dimensions. The initial state is a Gaussian spatial distribution of particles.

Since we have no *a priori* knowledge of the electronic dispersion of this system and the initial distribution of particles along this dispersion, we carry out MC simulations for a range of different dispersions and parameters in figs. S7 and S8. We begin by explaining the procedure with a simple linear dispersion model (all particles sharing the same speed) in fig. S7 before detailing the results of the other models.

To understand the relationship between the msd power law and the number of scattering events, the scattering time and microscopic particle velocity (slope of the linear dispersion) were systematically varied, generating images analogous to stroboSCAT for each simulation time step. We fit the simulated density profile using the same procedure as that used for stroboSCAT. Fig. S7A confirms that the average number of observed scattering events per particle over the observation time, $T_{obs}$, is set by the scattering time: $N_{\text{scattering}} = T_{obs}/\tau$. $N_{\text{scattering}}$ shows a measurable dependence on the velocity only for instances when a significant portion of particles escape the experimental window. Fig. S7B shows the extracted values of $\alpha$ (where msd $= At^\alpha$) over the same parameter range. $\alpha$ has a strong $\tau$ dependence and a weak dependence on the velocity, indicating that the curvature of the msd is a nearly independent gauge into the scattering time of a particle. Once the number of scattering events per particle becomes less than ~10 within $T_{\text{obs}}$, $\alpha$ becomes noticeably larger than 1. Fig. S7C shows the value of the msd at $t = 1.1$ ns as a function of the scattering time and particle velocity. Unlike $\alpha$, the msd strongly depends on both the scattering time and particle velocity.

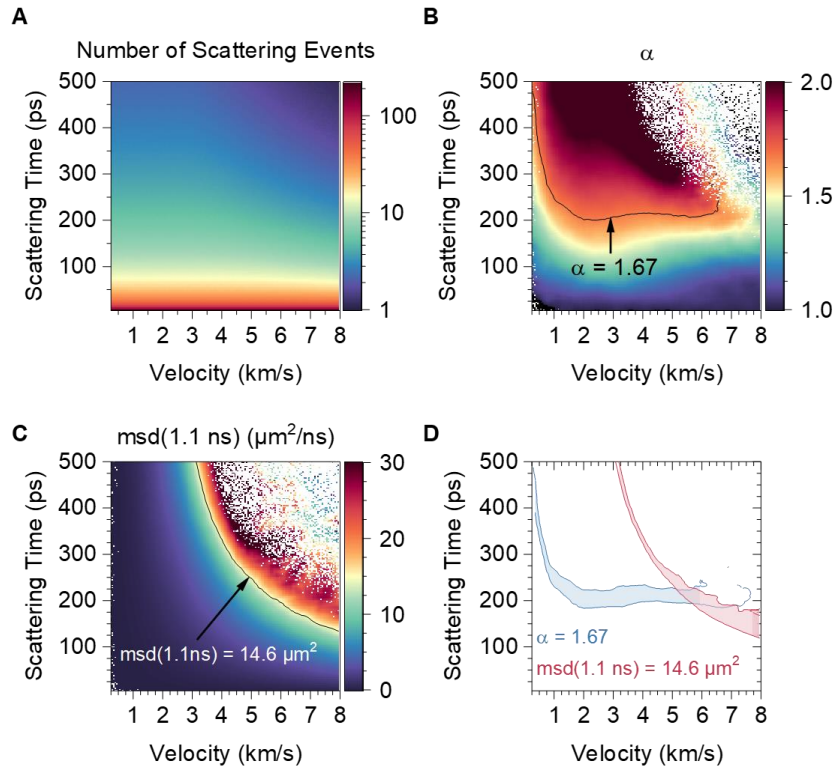

**Figure S7. Results of Monte Carlo simulations to model superdiffusive transport.** (A) The average number of scattering events per particle as a function of the scattering time and microscopic velocity for a total simulation time of 1.1 ns. (B) The exponent $\alpha$ resulting from a fit to msd $= At^\alpha$ of the simulated particle density profiles. (C) The msd value at 1.1 ns as a function of scattering time and particle velocity. White pixels in panels B and C are regions where Gaussian fits failed due to a significant number of particles exiting the simulation window. (D) Contours of $\alpha = 1.67$ (blue) and msd (1.1 ns) = 14.6 μm² (red). The crossing of these contours, at around $\tau = 215$ ps and $v = 5.5$ km/s, represents simulation parameters that reproduce the experimental msd. The shaded area refers to 95% confidence intervals in the fitting procedure.

To extract the simulation parameters that reproduce the experimental data shown in Figure 1G [$\alpha = 1.67$, msd (1.1 ns) $= 14.6$ μm²], we plot contours of constant $\alpha = 1.67$ and constant msd (1.1 ns) = 14.6 μm² extracted from panels B and C in fig. S7D. The 95% confidence intervals of the fit for each simulation is used to determine upper and lower bounds for the contours. The intersection of the $\alpha = 1.67$ and msd = 14.6 μm² contours are the parameters that reasonably reproduce the observed data. For this linear dispersion model, we extract $\tau = 215$ ps and $v = 5.5$ km/s, suggesting a mean free path of 1.2 μm.

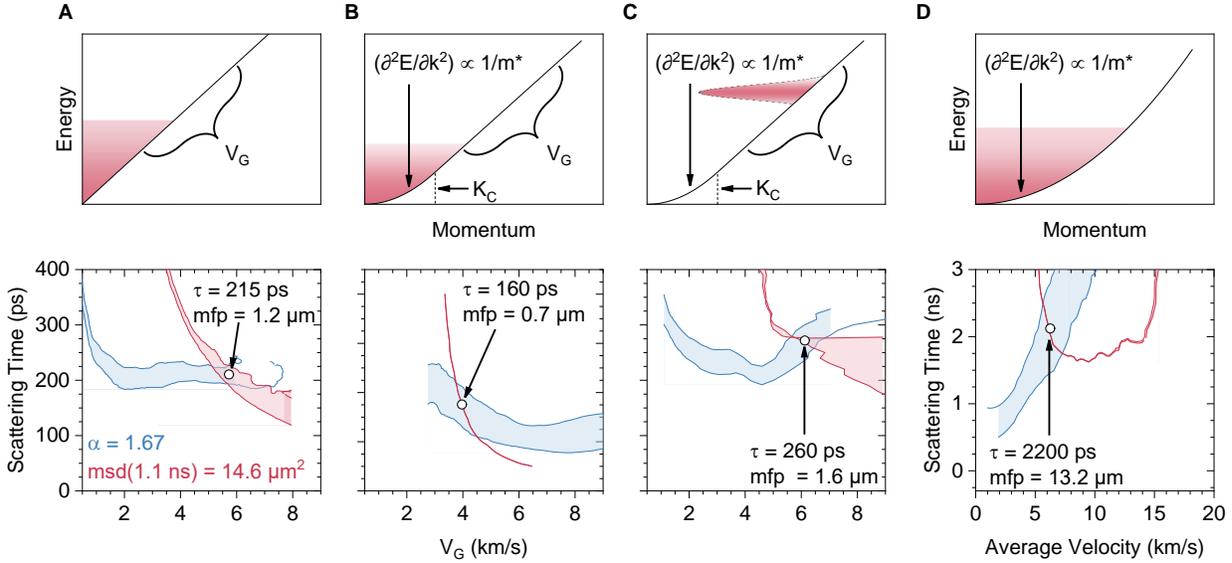

**Figure S8. Monte Carlo simulations for different model dispersions.** (A) Linear dispersion, equivalent to data shown in fig. S7. (B) Parabolic-linear dispersion consistent with our theory of acoustic polarons, starting with a thermal distribution. (C) Same as panel B, starting with a non-thermal distribution and incorporating inelastic scattering. (D) Parabolic dispersion, where the x-axis reports the average particle velocity for a given effective mass. Contours plotted in the bottom row are for $\alpha = 1.67$ (blue) and msd (1.1 ns) = 14.6 μm² (red). The shaded area refers to 95% confidence intervals in the fitting procedure. Intersections of the shaded areas reproduce the experimental msd. The white circles and associated labels are the central values of the areas.

Fig. S8 repeats the MC analysis for a range of model dispersions. Specifically, we investigate (top row, fig. S8): (i) A simple linear dispersion with all particles sharing the same input velocity dictated by the slope of the dispersion (fig. S8A); (ii) A parabolic-linear dispersion consistent with our theory of acoustic polarons in $Re_6Se_8Cl_2$, displayed in Fig. 4B of the main text. Two different simulations for this parabolic-linear dispersion are carried out: one with a thermalized (Boltzmann) starting condition (fig. S8B), and one with a non-equilibrium starting condition which thermalizes over time through inelastic scattering with acoustic phonons (fig. S8C). The parabolic-linear dispersion is constructed with a piecewise function composed of a parabola and a line at a crossover momentum ($K_C$). Enforcing continuity and differentiability at a fixed parabolic-linear crossover results in an effective mass, $m^*$, of the parabolic band that depends on the input particle velocity, $V_G$ (which changes the slope of the linear part of the dispersion). Note that implementing rigorous 1-phonon momentum-conserving scattering leads to zero phonon emission for both very heavy mass and linear dispersion (*53*, *54*), so that a thermal distribution is never established. Instead, we incorporate inelastic scattering in the nonequilibrium simulation phenomenologically through absorption or emission of acoustic phonons at a single frequency, ensuring energy but not momentum conservation. (iii) A parabolic dispersion with a thermalized distribution (fig. S8D). For the latter, we plot the average velocity of particles across the distribution.

The contours and resulting scattering times and velocities that reproduce the experimental msd are plotted in the bottom row of fig. S8. We find that the scattering times and mean free paths extracted for the different linear and linear-parabolic models are consistent, ranging from 160–260 ps and 0.7–1.6 µm respectively. The linear dispersion model falls in between the two linear-parabolic models, so we report numbers for the linear model in the main text. For the thermalized parabolic model, the scattering time necessary to reproduce the experimental msd implies ballistic transport over the full experimental window. The difficulty in attaining a superdiffusive msd for the parabolic model arises because a large effective mass implies a very broad distribution of velocities for a thermal population. Thus, even if the scattering time is very long, transport appears mostly diffusive because of the spread in particle velocities.

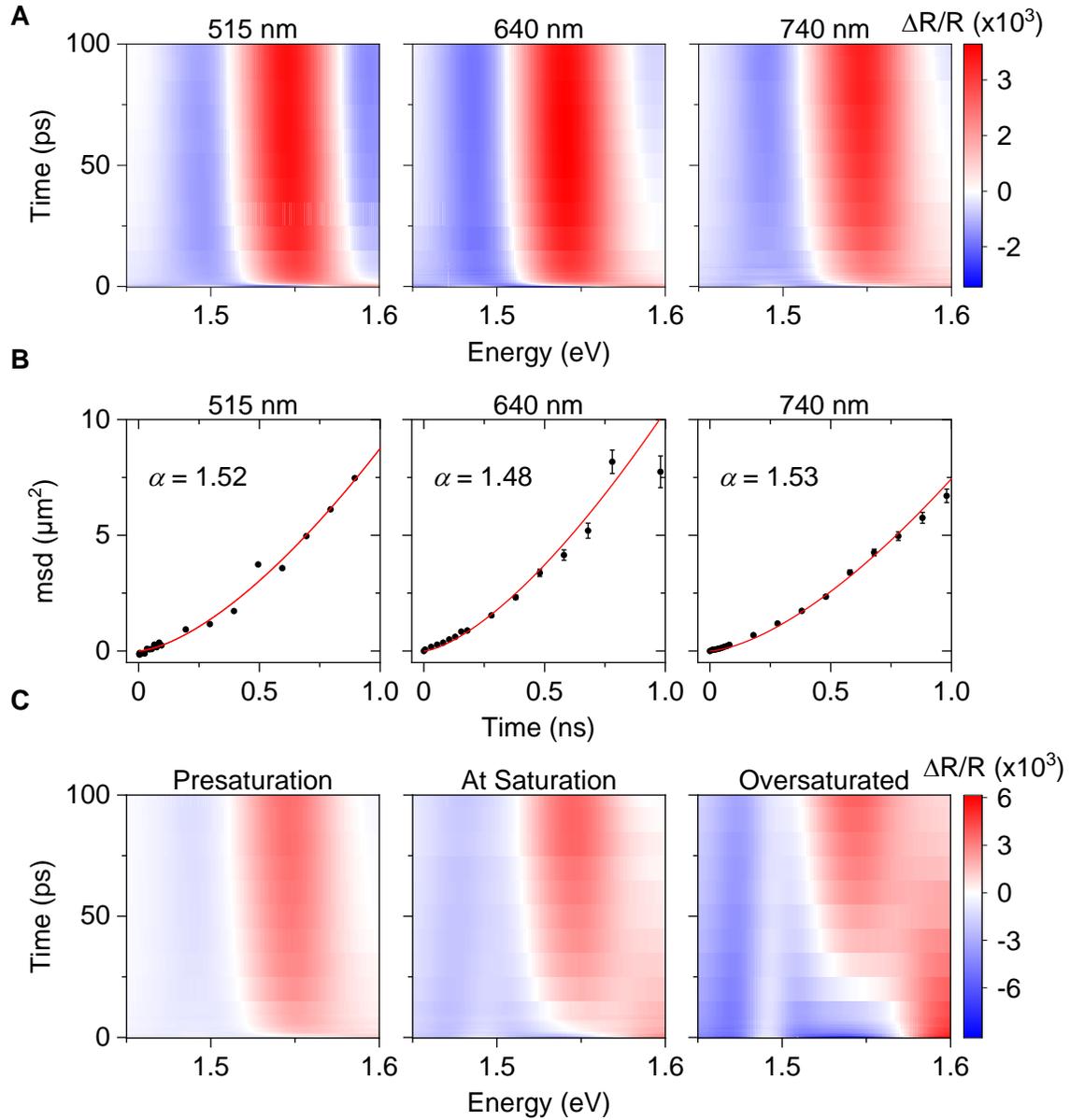

**Figure S9. Pump wavelength dependence on spectral and transport dynamics.** (A) Comparison of transient reflectance spectra using three pump wavelengths, 515 nm (used in the main text data), 640 nm and 740 nm at similar carrier densities of ~1.8 x $10^{18}$/cm$^3$. The spectral redshift dynamics are essentially identical. (B) Mean-square-displacement extracted from STR for different pump wavelengths show similar superdiffusive behavior. (C) Transient reflectance spectra at different pump fluences using a 740 nm excitation pulse reproduces the saturation effect observed with a 515 nm excitation pulse.

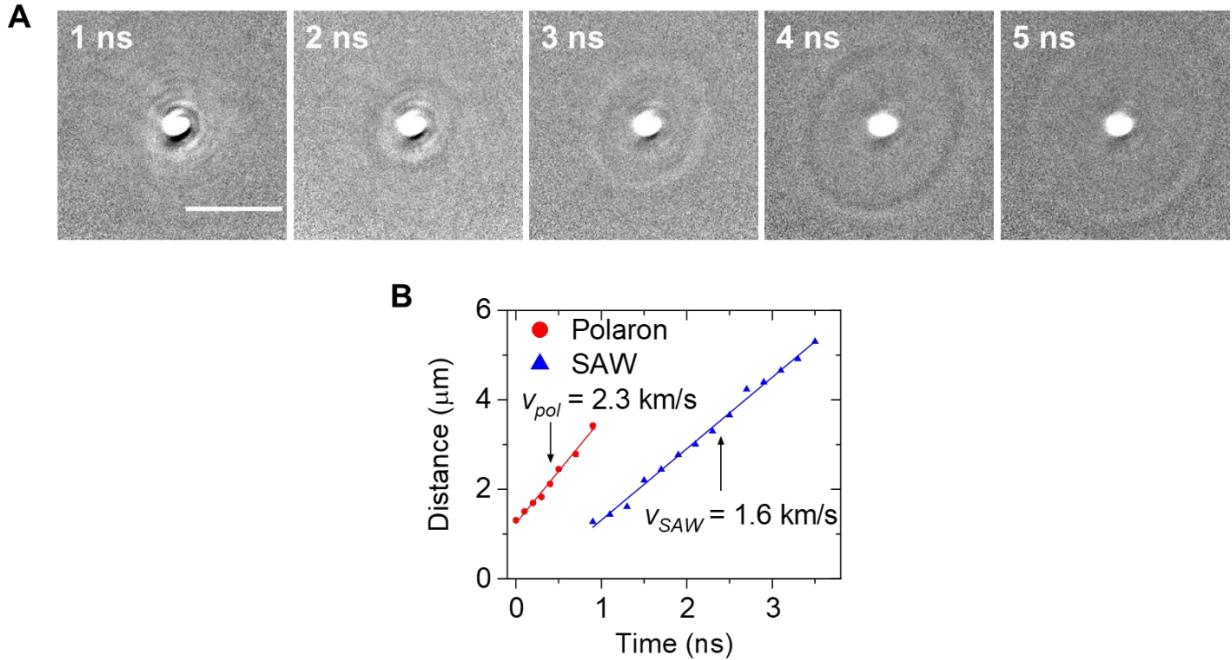

**Figure S10. Distinguishing surface acoustic waves (SAW) from polarons.** (A) Exciting $Re_6Se_8Cl_2$ with high pump densities (initial exciton density ~ $10^{20}$/cm$^3$) launches SAWs, which propagate ballistically (ring-like feature observed in stroboSCAT). Here, the stroboSCAT probe wavelength is set to 650 nm (1.91 eV) where polarons possess no spectral weight. The positive (white) signal in the center is due to the slow-moving excitons that persist due to saturation and have a spectral response at 650 nm (fig. S12). (B) A comparison of polaron effective velocity and the SAW velocity. Polaron transport precedes and far exceeds SAW propagation for similar time scales.

| Energy Carrier | Diffusivity/Effective Velocity |
| --- | --- |
| Acoustic exciton-polarons | 65 cm$^2$/s | 2300 m/s |
| Bare excitons (in saturation regime) | 0.008 cm$^2$/s |
| Surface Acoustic Waves | 1600 m/s |
| Intralayer heat | 0.0041 cm$^2$/s |
| Interlayer heat | 0.0007 cm$^2$/s |

**Table S2. Summary of different energy carriers in Re$_6$Se$_8$Cl$_2$.** Exciton diffusion can be measured in a saturated regime, off-resonant from the polaron energies (fig. S12). Surface acoustic waves at the interface of glass and crystal can be excited and measured under high fluence conditions (fig. S10). Thermal diffusivity, $D$, is calculated using $D = \kappa/\rho c_p$, where $\kappa$ is the thermal conductivity, $\rho$ is the mass density and $c_p$ is the specific heat capacity at constant pressure (J/gK). The thermal conductivity in and out of plane of the vdW layers is measured via Frequency Domain Thermoreflectance (*6*). The heat capacity is measured to be 0.213 J/gK at 298 K as discussed in supplementary section 2.3 (fig. S11).

2.3    Heat capacity measurement

The temperature-dependent specific heat capacity of Re$_6$Se$_8$Cl$_2$ was measured with a Quantum Design Physical Property Measurement System (PPMS) DynaCool, from 55 K to 300 K. To prepare the sample for measurement, powder Re$_6$Se$_8$Cl$_2$ was pressed into a pellet of mass 3.014 mg. The pellet was adhered to the Heat Capacity puck using Apiezon N Grease.

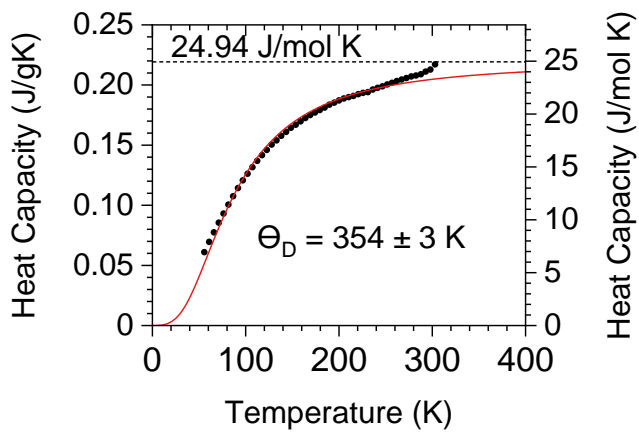

**Figure S11. Heat capacity.** The temperature-dependent specific heat capacity at constant pressure of Re$_6$Se$_8$Cl$_2$ was measured to estimate the thermal diffusivity in Table S1. The specific heat at 298 K is 0.213 J/gK, with no phase transition observed within the temperature range of 55 to 300 K. A Debye model is fit to estimate a Debye temperature of 354 K (red line). The dashed line corresponds to the classical limit. Note, per mole refers to moles of atoms, not clusters.

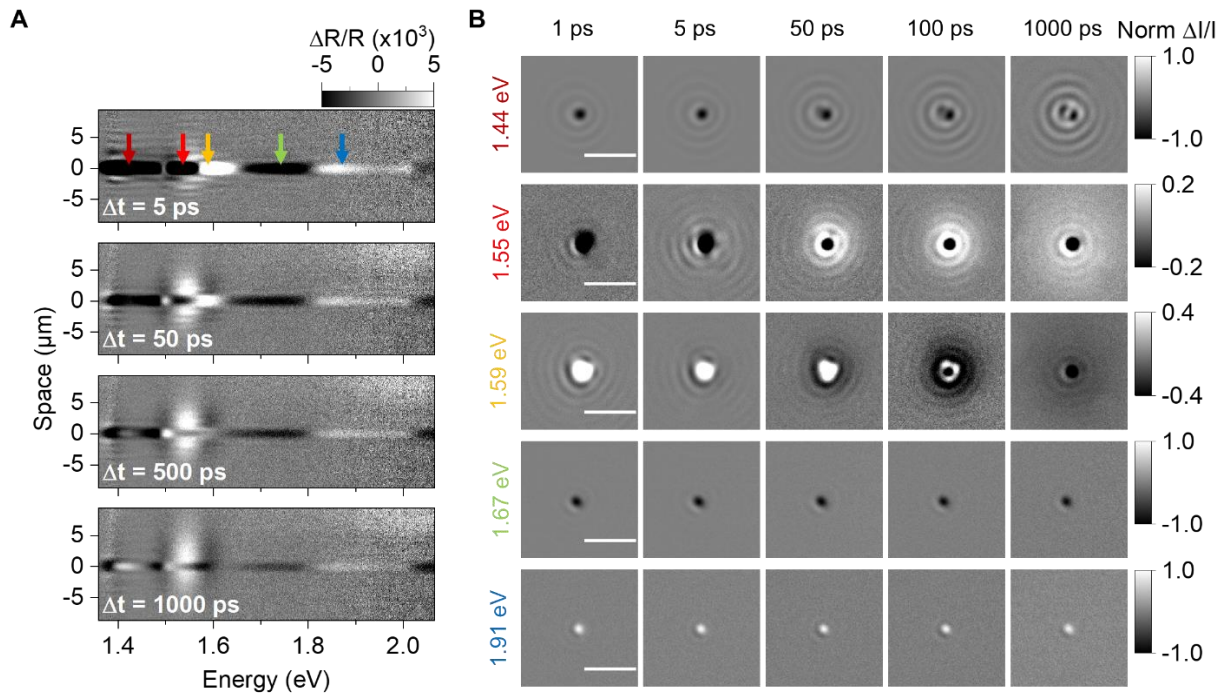

**Figure S12**. Correlated spatial and spectral dynamics. (A) STR for a few characteristic time delays taken at above-saturation fluences. (B) stroboSCAT images taken at different probe wavelengths. The colored arrows in panel A correspond to the energies reported in panel B. Probe energies at 1.55 eV (800 nm) and 1.59 eV (780 nm) show practically the same exciton and polaron density profiles but have opposite contrast. Using pump fluences above the polaron saturation limit allows us to visualize the exciton profile at energies that are spectrally isolated from the polaron, such as at 1.67 eV and 1.91 eV, thus providing an estimate of the bare exciton diffusivity (0.008 cm$^2$/s).

## 2.4 Polaron saturation densities estimated through population analysis

The relative exciton and polaron populations for Figure 3C are obtained through double-Gaussian fitting of the stroboSCAT population profiles as a function of pump fluence. The saturation trend is reproduced across a range of different experimental conditions such as pump area, flake location and pump wavelength. We found empirically that the evolution of the exciton population profile is most reliably fit when the pump excitation area is much larger than a diffraction-limited spot such that the polaron density profile evolution from a Gaussian (polarons only) to a super-Gaussian (flat-top Gaussian, representing polaron saturation in the center of the excitation region where the density is highest) to a double-Gaussian (polarons + unconverted excitons due to polaron saturation) is evident. Figure S13 displays this evolution as a function of initial peak excitation fluence at a pump-probe time delay of 5 ps. Using the peak fluence, we calculate along the y-axis an effective polaron density at each point in space. The density at which the flat-top Gaussian emerges thus provides a measure of the polaron saturation density. Beyond the saturation limit, unconverted excitons (negative contrast) emerge.

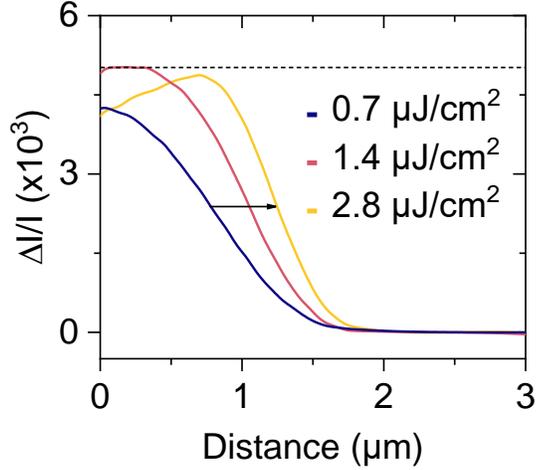

**Figure S13. Selected datasets of the density profile at increasing fluences.** From low to high fluences, the polaron amplitude increases until it reaches a maximum amplitude, generating a flat-top Gaussian. Subsequently, the amplitude decreases due to the overlapping exciton population (negative contrast).

To generate Figure 3C of the main text, the polaron density profile is modelled as

$$\rho_P(r, t; \lambda) = P_0 \varepsilon_P(\lambda) \exp\left(-\frac{(r-r_0)^p}{2\sigma^2}\right)$$

Where the only new parameter is $p$, which has a lower limit $p \geq 2$. When $p = 2$, the profile is Gaussian. $p > 2$ implies a flat-top Gaussian. $P_0$ is fixed to the maximum amplitude before the exciton population is appreciably visible, representing the polaron saturation density. Excitons follow a standard Gaussian profile of

$$\rho_X(r, t; \lambda) = X_0 \varepsilon_X(\lambda) \exp\left(-\frac{(r-r_0)^2}{2\sigma^2}\right)$$

The relative populations are determined via integration of the two functions in 2D polar coordinates. For the displayed plot, the populations are defined as $\frac{P_0}{2.18}\sigma_P^2 \Gamma\left(1 + \frac{2}{p}\right)$ and $X_0 \sigma_X^2$ for polarons and excitons respectively. The factor of 2.18 accounts for the difference in spectral amplitudes between polarons and excitons (this factor is used simply for display purposes and does not affect the calculated polaron saturation density), and $\Gamma$ is the Gamma function to account for radial integration of a super-Gaussian function.

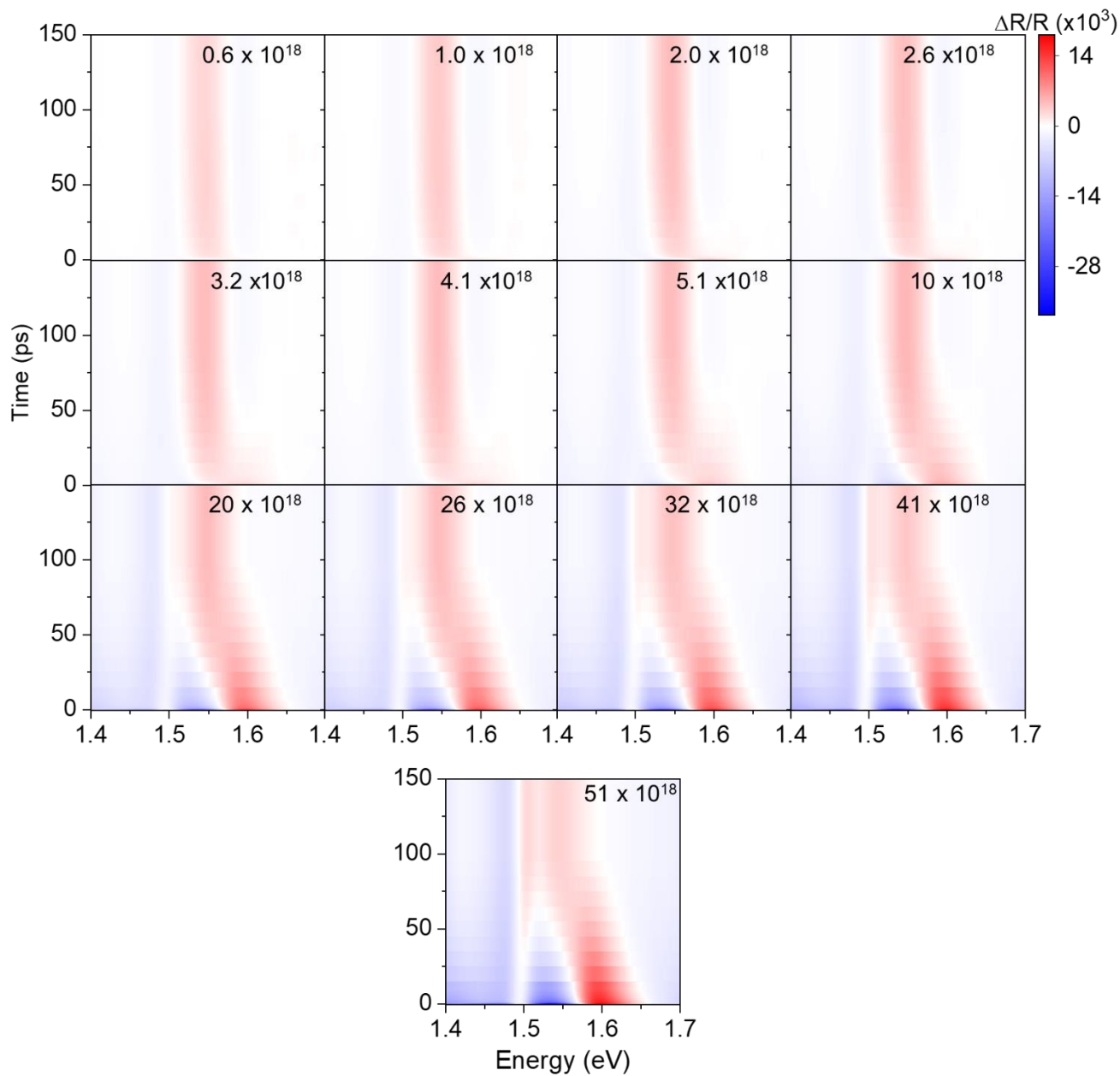

**Figure S14. Spectral evolution for spacetime transient reflection measurements as a function of initial carrier density**. All numbers are peak carrier densities reported in /cm$^3$. As discussed in the main text, polaron saturation results in a prolonged redshift associated with exciton to polaron conversion from a few picoseconds to tens of picoseconds.

## 2.5 STR Simulation

### 2.5.1 Spectral analysis

Figure S15 illustrates the procedure used to extract bare exciton and polaron spectra from STR; these spectral shapes are used as inputs in the simulations for figure 3E. The raw spacetime transient reflectance is sliced along two spatial axes (fig. S15A): one in the center of the excitation spot and another off-axis. The former shows the spectral evolution from excitons to polarons; the off-axis cut displays the pure polaron spectrum since only polarons can propagate over that distance. A 4-component sequential global analysis model of A→B→C→D is used to model the central spectra as shown in figure S15B. The first spectral component "A" represents the species prior to both propagation and polaron conversion and is thus taken to represent the exciton spectrum. The polaron spectrum is extracted from an off-axis component of STR as shown in figure S15C.

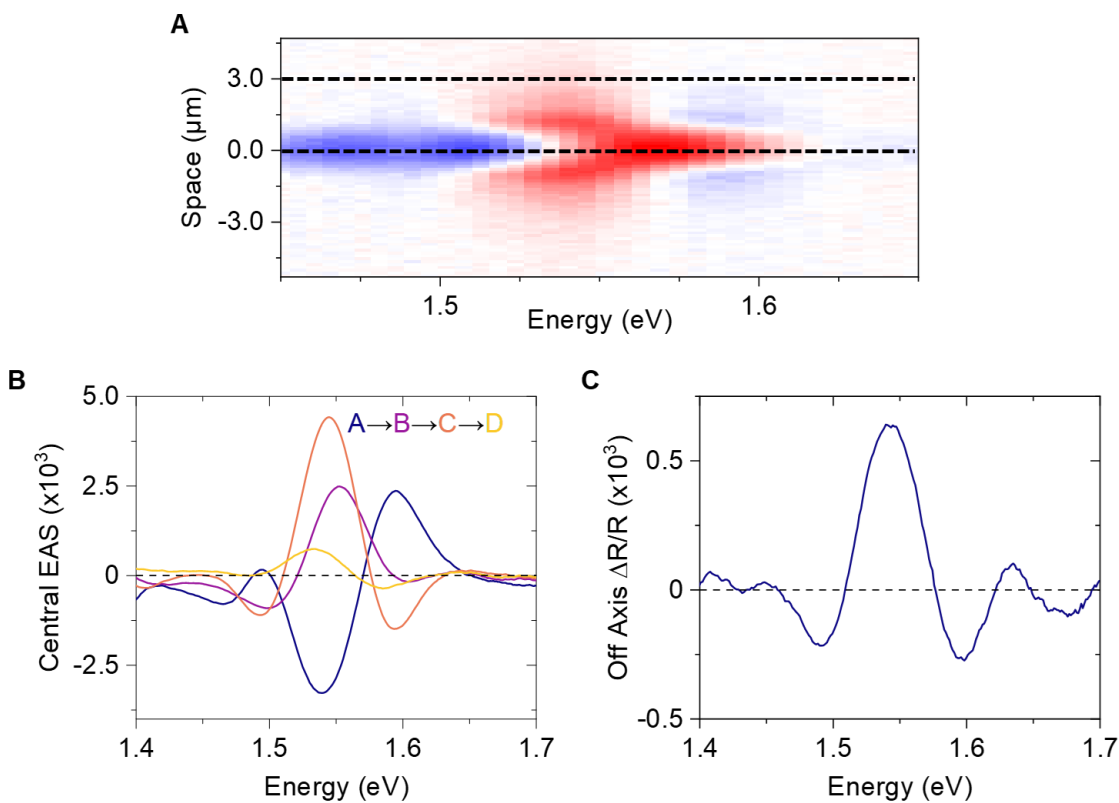

**Figure S15. Determination of exciton and polaron spectra for spectral simulations.** (A) A raw image of spacetime TR for 100 ps (taken from the main text). (B) Evolution associated spectra (EAS) resulting from a 4-component sequential kinetic model with lifetimes of 0.8 ps, 36 ps, 400 ps and 4000 ps (note that these lifetimes represent both population decay and 2D propagation out of the excited and probed regions). (C) The off-axis pure polaron spectrum. Component "C" is the same as the off-axis trace and is therefore taken to represent the polaron spectrum.

2.5.2  Simulation of density dependent transient reflectance

A phenomenological model is constructed to simulate the density-dependent kinetics of an exciton to polaron transformation as observed shown in figure 3E of the main text. To account for decay due to transport, a treatment beginning with the continuity equation is required. Three simplifying assumptions are made for the purposes of the model:

1. The exciton to polaron transformation, $X \to P$, is first order and proceeds under a rate $k = k[P(r,t)]$, where the rate has an implicit dependence on radius and time through the polaron number, $P(r,t)$. For the simulation, the rate is phenomenologically modelled as $k[P(r,t)] = k_{micro}(1 - P(r,t)/N)$ where $k_{micro}$ is the rate of the exciton to polaron transition at below-saturation fluences, and $N$ is the number of available sites for polarons to form. Thus, $k = 0$ when the polaron sites are totally filled. $N$ is estimated as the ratio of the excitation area to the polaron area i.e. polarons cannot overlap in space. This estimate relies on the assumption that excitons do not diffuse substantially out of the excitation area within the first ~100 ps. This assumption is justified given the experimentally inferred rate of 0.008 cm$^2$/s for bare exciton diffusion in Re$_6$Se$_8$Cl$_2$ (Table S2, fig. S12).

2. The particle flux is assumed to be diffusive and isotropic in 2-dimensions, $\vec{j}_P = -D_P \vec{\nabla} P$ which neglects the complexity of density-dependent scattering. The excitons are assumed to be motionless, $\vec{j}_X = 0$. Although we use a diffusive (rather than superdiffusive) model for simplicity, we note that an equivalent treatment using ballistic flux ($\vec{j}_P = v\,\hat{r}$) results in almost identical spectral evolution.

3. Polaron decay is neglected due to long polaron lifetime of ~11 ns (fig. S5).

With the above assumptions, the following equations are numerically solved for $X(r,t)$ and $P(r,t)$:

$$\frac{\partial X}{\partial t} + \nabla \cdot \vec{j}_X = \frac{\partial X}{\partial t} = -k[P(r,t)]X(r,t)$$

$$\frac{\partial P}{\partial t} + \nabla \cdot \vec{j}_P = \frac{\partial P}{\partial t} - D_P \nabla^2 P = k[P(r,t)]\,X(r,t)$$

$$X(r, t=0) = X_0 e^{-\frac{r^2}{2\sigma^2}}$$

$$P(r, t=0) = 0$$

The simulation requires spectral traces that represent the initial excitonic state and final polaronic state, $\varepsilon_X(\lambda), \varepsilon_P(\lambda)$ respectively, where $\lambda$ is the wavelength. The simulated data is calculated as

$$\varepsilon_P(\lambda)P(r,t) + \varepsilon_X(\lambda)X(r,t)$$

Where $\varepsilon_X(\lambda)$ and $\varepsilon_P(\lambda)$ are extracted using the method detailed in section 2.5.1 and figure S15 above.

The simulations use empirically determined values for effective diffusion and polaron radius, and a microscopic rate constant $k_{micro}= (1.5 \text{ ps})^{-1}$ which averages over the complexity of the damped transition shown in Figure 2B of the main text. Using these parameters, we then vary the initial exciton population. The saturation trend can be reproduced as shown in Figure 3E of the main text.

## 2.6 Theoretical and computational methods

### 2.6.1 Density functional theory calculations

We performed density functional theory (DFT) calculations of $Re_6Se_8Cl_2$ with Quantum Espresso (*55*, *56*), using the PBE exchange correlation functional, projector augmented wave pseudopotentials, a kinetic energy cutoff of 60 Ry, and a 5x5x1 *k*-point mesh to sample the first Brillouin zone. We optimized the internal geometry of the atoms while keeping the lattice parameters fixed to their experimental values. The calculated band structure is shown in Figure 1B.

### 2.6.2 Exciton effective mass

We extract the electron and hole effective masses by fitting the band structure of Figure 1B at the conduction band minimum and valence band maximum to a parabolic form. We perform this fit along the high-symmetry directions: R–T–Z for the electron mass $m_e^*$ and X–Γ–Y for the hole mass $m_h^*$. The effective mass of the exciton is then calculated to be $m = m_e^* + m_h^* = 1.88\, m_e$, where $m_e$ is the fundamental mass of the electron.

### 2.6.3 Deformation potential

To estimate the in-plane coupling strength between excitons and acoustic phonons, we calculate the uniaxial deformation potential, $D = \Delta E_g/\epsilon$, where $\Delta E_g$ is the change in the band gap due to a small strain $\epsilon$ (*57–59*). We simulate this strain by modifying the lattice parameter $a$ to be 1% larger and smaller and relaxing the atomic positions. The band gap increases by 50.3 meV and decreases by 38.6 meV with 1% tensile and compressive strain which gives an average deformation potential of $D = 4.4$ eV. This calculation neglects excitonic effects that might modify the deformation potential $D$.

### 2.6.4 Speed of sound

We estimate the in-plane speed of sound using elastic constants calculated from DFT and the 2D mass density $\rho$ (*60*, *61*). We first optimize the cell parameters and then calculate the total energy (allowing atomic relaxation) with strain $\epsilon_x$ and $\epsilon_y$ in the *x* and *y* directions, each varying from -2% to 2%. We then fit the total energy to the form $E = E_0 + a_{xx}\epsilon_x^2 + a_{yy}\epsilon_y^2 + a_{xy}\epsilon_x\epsilon_y$, allowing the calculation of the in-plane stiffness $C$ and Poisson's ratio $v$,

$$C = \frac{1}{A}\left(2a_{xx} - \frac{a_{xy}^2}{2a_{xx}}\right)$$

$$v = \frac{a_{xy}}{2a_{xx}}$$

where $A$ is the area and we assumed the material to be homogeneous and isotropic in 2D ($a_{xx} = a_{yy}$). The longitudinal speed of sound is then calculated to be

$$s = \sqrt{\frac{C(1-v)}{\rho(1+v)(1-2v)}} \approx 5 \text{ km/s}$$

### 2.6.5 Renormalization of the exciton transfer integral

To calculate the exciton transfer integral ($J$) at room temperature, we combine the calculated zero-temperature $J_0$ with the experimentally-observed temperature-dependent bandwidth narrowing from angle-resolved photoemission spectroscopy measurements of this material (7),

$$J(T) = J_0 \exp[-g^2(n_\omega + 1/2)]$$

where $g = 1.2$ and the Bose-Einstein occupancy is $n_\omega = 1.9$ for optical phonons of energy 10.8 meV at 300 K. This procedure yields $J(T = 300 \text{ K}) = 1.5$ meV. This renormalization approximately accounts for local coupling to high-frequency optical phonons that are otherwise neglected in our theory (62).

| Material | Unit cell mass (amu) | Deformation potential (eV) | Speed of sound (km/s) | Exciton Transfer Integral[a] (meV) |
|---|---|---|---|---|
| Re$_6$Se$_8$Cl$_2$ | 1820 | 4.4[b] | 5[c] | 1.5 |
| Pentacene | 556[d] | 1.74 (63) | 2.8 (64) | 30 |
| WSe$_2$ | 342 | 7[e] (59) | 5 (65) | 480 |
| 2D halide perovskites[f] | 837 | 0.9 (66) | 3 (67) | 113 (68) |

[a] Section 2.6.5; [b] Section 2.6.3; [c] Section 2.6.4
[d] Pentacene has 2 molecules per unit cell.
[e] Value taken to be calculated deformation potential of conduction band minimum at K point in Brillouin zone.
[f] (PEA)$_2$PbI$_4$ parameters are used

**Table S3. Comparison of values for coupling strength $\lambda$.**

### 2.6.6 Acoustic polaron binding energy and dispersion

Assuming a Frenkel-like (molecular) exciton description, we calculate the quantum mechanical properties of the acoustic exciton-polaron using a two-dimensional Su-Schrieffer-Heeger (SSH) Hamiltonian with periodic boundary conditions describing an exciton on a lattice with nonlocal coupling to acoustic phonons (*69*). The parameters were chosen according to the calculations described above (for an SSH exciton-phonon coupling of the form $\alpha \sum \hat{a}_n^\dagger \hat{a}_m (\hat{x}_n - \hat{x}_m)$, the SSH exciton-phonon coupling strength is related to the deformation potential by $\alpha = D/2a$). The polaron binding energy and dispersion were calculated using the Davydov-like variational wavefunction

$$|\Psi\rangle = |\Psi_{ex}\rangle |\Psi_{ph}\rangle$$
$$|\Psi_{ex}\rangle = \sum_n e^{ipn} \psi_n \hat{a}_n^\dagger |0\rangle_{ex}$$
$$|\Psi_{ph}\rangle = \exp\left[\sum_n d_n (\hat{b}_n - \hat{b}_n^\dagger)\right] |0\rangle_{ph}$$

where $\psi_n$ and $d_n$ are the exciton wavefunction and lattice displacement, respectively. Although this wavefunction is not an eigenstate of the lattice translation operator, the dispersion relation was obtained by minimizing the energy over the variational parameters $p$, $\psi_n$, and $d_n$, while constrained to have a fixed expectation value of the total exciton + phonon momentum (*39*). Calculations were performed on a periodic lattice of 16x16 sites.

At zero momentum, this theory predicts that the polaron is unbound below the same critical coupling strength as the one derived in the main text. An improved wavefunction will be translational symmetry adapted with more variational parameters. From such a theory we expect a binding energy that is larger in magnitude and nonzero for all values of the coupling strength (*70*). In this light, the critical value of the coupling strength used in this work ($\lambda_c = 1.6$) should be understood as an estimate of the crossover between strongly bound and weakly bound at zero temperature. A more detailed theoretical study will be presented elsewhere.

### 2.7 Temperature-dependent stroboSCAT

The mean squared displacements obtained from temperature-dependent stroboSCAT in fig. S16A show that fast and long-range transport is only observed at temperatures greater than 80 K. We plot the value of the msd at 800 ps to provide a model-independent handle on transport properties. At low temperatures, excitons exhibit slow diffusive behavior. For example, $D = 0.15$ cm$^2$/s at 10 K, displaying a msd value at 800 ps that is 2 orders of magnitude lower than at room temperature. The msd rises rapidly over the range 80–125 K and peaks at 200 K. We assign this transition from slow to fast transport to polaron binding, in good agreement with our theory which predicts that polarons are stable above 175 K (fig. S16B). The decrease in msd observed above 200 K is tentatively assigned to weak polaron–phonon scattering, which decreases the mean free path of polarons as temperature is increased.

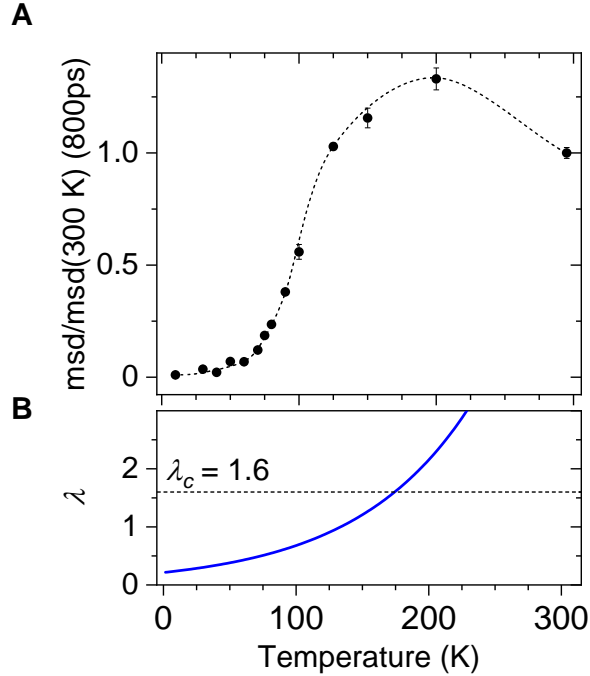

**Figure S16**. **Temperature-dependent stroboSCAT experiments on Re$_6$Se$_8$Cl$_2$** (A). The ratio of temperature dependent msd at 800 ps to the msd at room temperature at 800 ps. The value of the msd at 800 ps is 2 orders of magnitude lower at 10 K than at room temperature. The dashed line is a spline to guide the eye. All experiments are performed using 515 nm excitation. (B) Value of $\lambda$ as a function of temperature for Re$_6$Se$_8$Cl$_2$. Here the temperature dependence stems from $\lambda \propto 1/J$, and $J(T) = J_0 \exp[-g^2(n_\omega + 1/2)]$ (section 2.6.5). The crossing point $\lambda = \lambda_c$ occurs at 175 K, below which the acoustic polaron is predicted to unbind.